\documentclass[aip,jcp,reprint,noshowkeys]{revtex4-2}
\usepackage{graphicx,dcolumn,bm,xcolor,microtype,multirow,amscd,amsmath,amssymb,amsfonts,physics,wrapfig,bbold,siunitx,xspace}
\usepackage[version=4]{mhchem}

\usepackage[utf8]{inputenc}
\usepackage[T1]{fontenc}

\usepackage{hyperref}
\hypersetup{
    colorlinks,
    linkcolor={red!50!black},
    citecolor={red!70!black},
    urlcolor={red!80!black}
}

\usepackage{listings}
\definecolor{codegreen}{rgb}{0.58,0.4,0.2}
\definecolor{codegray}{rgb}{0.5,0.5,0.5}
\definecolor{codepurple}{rgb}{0.25,0.35,0.55}
\definecolor{codeblue}{rgb}{0.30,0.60,0.8}
\definecolor{backcolour}{rgb}{0.98,0.98,0.98}
\definecolor{mygray}{rgb}{0.5,0.5,0.5}

\definecolor{sqred}{rgb}{0.85,0.1,0.1}
\definecolor{sqgreen}{rgb}{0.25,0.65,0.15}
\definecolor{sqorange}{rgb}{0.90,0.50,0.15}
\definecolor{sqblue}{rgb}{0.10,0.3,0.60}

\lstdefinestyle{mystyle}{
    backgroundcolor=\color{backcolour},
    commentstyle=\color{codegreen},
    keywordstyle=\color{codeblue},
    numberstyle=\tiny\color{codegray},
    stringstyle=\color{codepurple},
    basicstyle=\ttfamily\footnotesize,
    breakatwhitespace=false,
    breaklines=true,
    captionpos=b,
    keepspaces=true,
    numbers=left,
    numbersep=5pt,
    numberstyle=\ttfamily\tiny\color{mygray},
    showspaces=false,
    showstringspaces=false,
    showtabs=false,
    tabsize=2
  }

  \newcolumntype{d}{D{.}{.}{-1}}

\newcommand{\dwar}{\downarrow}
\newcommand{\upar}{\uparrow}

\newcommand{\rs}{r_s}

\newcommand{\rhos}{\rho_\sigma}
\newcommand{\rhoup}{\rho_\upar}
\newcommand{\rhodw}{\rho_\dwar}
\newcommand{\eps}{\epsilon}
\newcommand{\br}{\boldsymbol{r}}
\newcommand{\bp}{\boldsymbol{p}}
\newcommand{\bq}{\boldsymbol{q}}
\newcommand{\bk}{\boldsymbol{k}}
\newcommand{\kF}{k_\text{F}}
\newcommand{\kFs}{k_{\text{F}\sigma}}
\newcommand{\kFup}{k_{\text{F}\upar}}
\newcommand{\kFdw}{k_{\text{F}\dwar}}
\newcommand{\Dels}{\Delta_{\sigma}}

\newcommand{\kaps}{\kappa_{\sigma}}

\newcommand{\titou}[1]{\textcolor{black}{#1}}

\newcommand{\LCPQ}{Laboratoire de Chimie et Physique Quantiques (UMR 5626), Universit\'e de Toulouse, CNRS, France}

\begin{document}	

\title{Excited States of the Uniform Electron Gas}
\author{Pierre-Fran\c{c}ois \surname{Loos}}
	\email{loos@irsamc.ups-tlse.fr}
	\affiliation{\LCPQ}
	
\begin{abstract}
The uniform electron gas (UEG) is a cornerstone of density-functional theory (DFT) and the foundation of the local-density approximation (LDA), one of the most successful approximations in DFT.
In this work, we extend the concept of UEG by introducing excited-state UEGs, systems characterized by a gap at the Fermi surface created by the excitation of electrons near the Fermi level.
We report closed-form expressions of the reduced kinetic and exchange energies of these excited-state UEGs as functions of the density and the gap. 
Additionally, we derive the leading term of the correlation energy in the high-density limit.
By incorporating an additional variable representing the degree of excitation into the UEG paradigm, the present work introduces a new framework for constructing local and semi-local state-specific functionals for excited states.
\end{abstract}

\maketitle

\section{Introduction}
\label{sec:introduction}

The development of state-specific functionals for excited states \cite{PerLev-PRB-85,Gor-PRA-96,Gor-PRA-99,LevNag-PRL-99,AyeLev-PRA-09,AyeLevNag-PRA-12,AyeLevNag-JCP-15,AyeLevNag-TCA-18,Gar-ARMA-22,GiaLoo-JPCL-23,GouPit-PRX-24,LooGia-JCP-25,GouDalKroPit-arXiv-24,Gou-PRA-25,YanAye-arXiv-24,Fro-JPCA-25} marks a pivotal advancement in density-functional theory (DFT). \cite{HohKoh-PR-64,KohSha-PRA-65,TeaHelSav-PCCP-22}
While the local-density approximation (LDA) \cite{Tho-MPCPS-27,Fer-RANL-27,Dir-PCPS-30,CepAld-PRL-80,LewLieSei-PAA-20} often serves as the foundational starting point for constructing functionals, \cite{Sla-PR-51,VosWilNus-CJP-80,PerZun-PRB-81,PerWan-PRB-92,Cha-JCP-16} a critical question remains: can we design local and/or semi-local functionals --- one that depends solely on local variables such as the electron density --- tailored for electronic excited states?

A significant challenge lies in the lack of an established paradigm to construct excited-state uniform electron gases (UEGs). 
While the ground-state UEG, \cite{ParrBook,VignaleBook,LooGil-JCP-11,LooGil-WIREs-16,LewLieSei-JEP-18} also known as the homogeneous electron gas or jellium model, is well-understood and serves as the foundation of most existing local approximations, their excited-state counterparts are far less straightforward. 
By modifying the UEG paradigm to account for electronic excitations, we aim to create a framework capable of describing excited states in a state-specific manner. \cite{PerLev-PRB-85,Gor-PRA-96,Gor-PRA-99,LevNag-PRL-99,AyeLev-PRA-09,AyeLevNag-PRA-12,AyeLevNag-JCP-15,AyeLevNag-TCA-18,Gar-ARMA-22,GiaLoo-JPCL-23,GouPit-PRX-24,LooGia-JCP-25,GouDalKroPit-arXiv-24,Gou-PRA-25,YanAye-arXiv-24,Fro-JPCA-25}
To achieve this, we propose introducing an additional local variable that quantifies the ``degree of excitation'' of the system. 
This variable would encode information about the nature and extent of electronic excitations, allowing the functional to adapt to the specifics of the excited states.

To address this, we propose a model for excited-state UEGs constructed by introducing a gap at the Fermi surface through excitations of the electrons near the Fermi level. 
These excited-state UEGs can be viewed as a generalization of the ``jellium with a gap'' model, where unoccupied states are rigidly shifted. \cite{Cal-PR-59,ReySav-IJQC-98,KriCheIafSav-99,GutSavKriChe-IJQC-99,KriCheKur-ACP-01,GutSav-PRA-07}
In such systems, while only the correlation functional is affected, \cite{Cal-PR-59,ReySav-IJQC-98,KriCheIafSav-99,KriCheKur-ACP-01,TreTerConOleSal-PRB-13,FabTreTerCon-JCTC-14,ConFabSmiDel-PRB-17,ConFabDel-PRB-18,JanConSam-JCP-23} both exchange and correlation functionals are influenced by the emergence of a gap in excited-state UEGs.
This provides a promising foundation for extending the LDA to incorporate excited states.
A related, albeit distinct, approach was explored by Harbola and collaborators, who focused on constructing LDA exchange functionals for excited states. \cite{SamHar-JPB-05,RahGanSamHarSahMoo-PhysicaB-09,HemHar-JMS-10,HemHar-EPJD-12,HemShaHar-JPB-14} 
However, their strategy clearly lacks generality and is restricted to exchange.
Alternative schemes have also been explored. \cite{Koh-PRA-86,ThePap-PRA-00,LooFro-JCP-20,MarSenFroLoo-FD-20}

Our approach builds on the recent work of Gould and Pittalis, who proposed incorporating excited-state information via the so-called constant occupation factor ensemble (cofe) UEGs. \cite{GouPit-PRX-24}
Here, we move beyond the ensemble framework to focus on a pure-state formalism.  
We believe this shift offers greater flexibility to capture the unique features of state-specific electronic excitations.
By embedding excitation-specific information into the functional, we hope to achieve a more accurate description of excited-state energetics and properties, paving the way for broader applications in quantum chemistry and materials science.
Atomic units are used throughout.

\section{Ground-State UEG}
\label{sec:GS}

The reduced (i.e., per electron) energy of the UEG is expressed as \cite{ParrBook,VignaleBook,LooGil-WIREs-16,LewLieSei-JEP-18}
\begin{equation}
	\eps(\rho,\zeta) = t_\text{s}(\rho,\zeta) + \eps_\text{x}(\rho,\zeta) + \eps_\text{c}(\rho,\zeta)
\end{equation}
where $t_\text{s}$, $\eps_\text{x}$, and $\eps_\text{c}$ are the kinetic, exchange and correlation energy components, $\rho = \rhoup + \rhodw$ is the (uniform) electron density, $\rhos$ is the density of the spin-$\sigma$ electrons ($\sigma$ = $\upar$ or $\dwar$), and the spin polarization is $\zeta = (\rhoup - \rhodw)/\rho$.
The Hartree contribution does not appear in the previous equation as it is exactly canceled by the uniform positively charged background.

Both the kinetic and exchange energies can be written as $t_\text{s}(\rho,\zeta) = t_\text{s}(\rho) \Upsilon_\text{s}(\zeta)$ and $\eps_\text{x}(\rho,\zeta) = \eps_\text{x}(\rho) \Upsilon_\text{x}(\zeta)$, where $t_\text{s}(\rho) \equiv t_\text{s}(\rho,\zeta=0)$ and $\eps_\text{x}(\rho) \equiv \eps_\text{x}(\rho,\zeta=0)$ are spin-unpolarized (or paramagnetic) quantities while the kinetic and exchange spin scaling functions are
\begin{subequations}
\begin{align}
	\Upsilon_\text{s}(\zeta) & = \frac{t_\text{s}(\rho,\zeta)}{t_\text{s}(\rho)} = \frac{(1-\zeta)^{5/3} + (1+\zeta)^{5/3}}{2}
	\\
	\Upsilon_\text{x}(\zeta) & = \frac{\eps_\text{x}(\rho,\zeta)}{\eps_\text{x}(\rho)} = \frac{(1-\zeta)^{4/3} + (1+\zeta)^{4/3}}{2}
\end{align}
\end{subequations}
The kinetic and exchange energies can also be spin-resolved as follows:
\begin{subequations}
\begin{align}
	t_\text{s}(\rho,\zeta) & = t_{\text{s}\upar}(\rhoup) + t_{\text{s}\dwar}(\rhodw)
	\\
	\eps_\text{x}(\rho,\zeta) & = \eps_{\text{x}\upar}(\rhoup) + \eps_{\text{x}\dwar}(\rhodw)
\end{align}
\end{subequations}
The spin-resolved correlation energy has an additional component and reads
\begin{align}
	\eps_\text{c}(\rho,\zeta) = \eps_{\text{c}\upar\upar}(\rhoup) + \eps_{\text{c}\dwar\dwar}(\rhodw) + \eps_{\text{c}\upar\dwar}(\rhoup,\rhodw)
\end{align}

In the usual ground-state spin-polarized (or ferromagnetic) UEG, the reduced kinetic energy is
\begin{equation}
	t_{\text{s}\sigma}(\rhos) 
	= \frac{1}{\rhos} \int_{0}^{\kFs} \frac{k^2}{2} \frac{k^2}{2\pi^2}\dd{k}
	= C_\text{F} \rhos^{2/3}
\end{equation}
where $C_\text{F} = - \frac{3}{10} \qty(6\pi^2)^{2/3} \approx 4.5578$ is the Thomas-Fermi coefficient \cite{Tho-MPCPS-27,Fer-RANL-27} and the uniform spin-density is
\begin{equation} \label{eq:rho}
	\rhos = \int_{0}^{\kFs} \frac{k^2}{2\pi^2}\dd{k} = \frac{\kFs^3}{6\pi^2}
\end{equation}
where $\kFs$ is the Fermi wave vector associated with the spin-$\sigma$ channel.
The exchange energy is given by the well-known Dirac formula \cite{Dir-PCPS-30,Fri-CMP-97} which reads
\begin{equation} \label{eq:ex}
\begin{split}
	\eps_{\text{x}\sigma}(\rhos) 
	& = \frac{1}{2} \iint \frac{\rho_\text{x}(\br_1,\br_2)}{r_{12}} \dd{\br_1} \dd{\br_2}
	\\
	& = C_\text{x} \rhos^{1/3}
\end{split}
\end{equation}
where $r_{12} = \abs{\br_1 - \br_2}$ is the interelectronic distance and $C_\text{x} = -\frac{3}{4} \qty(\frac{6}{\pi})^{1/3} \approx -0.930526$ is the Dirac coefficient.
In Eq.~\eqref{eq:ex}, $\rho_\text{x}(\br_1,\br_2) = - \abs{\rho_1(\br_1,\br_2)}^2/\rho(\br_1)$ is the Fermi hole fulfilling the normalization condition $\iint \rho_\text{x}(\br_1,\br_2) \dd{\br_1} \dd{\br_2} = -1$ and $\rho_1(\br_1,\br_2) = j_{\kFs}(r_{12})$ is the one-electron reduced density matrix with $j_{\kFs}(r_{12}) = 1/(2\pi^2) \qty[\sin(\kFs r_{12}) - \kFs r_{12} \cos(\kFs r_{12})]/(r_{12}^3)$.

Concerning the correlation part, one usually relies on perturbative expansions in the high- and low-density limits \cite{VignaleBook,LooGil-WIREs-16}.
As a function of the Wigner-Seitz radius $\rs = (\frac{3}{4\pi\rho})^{1/3}$, the small-$\rs$ (or high-density) expansion of the correlation energy appears to be \cite{Mac-ZNA-50,GelBru-PR-57,Dub-AP-59a,Dub-AP-59b,CarMar-PR-64,Mis-PR-65,OnsMitSte-AP-66,WanPer-PRB-91,Hof-PRB-92,EndHorTakYas-PRB-99,LooGil-PRB-11,LooGil-WIREs-16}
\begin{equation}
	\eps_\text{c}(\rs,\zeta) = \lambda_0(\zeta) \ln \rs + \eps_0(\zeta) + \lambda_1(\zeta) \rs \ln \rs + \eps_1(\zeta) \rs + \cdots
\end{equation}
where, as first proposed by Gell-Mann and Brueckner \cite{GelBru-PR-57}, one must rely on resummation techniques to avoid divergences.
This explains the appearance of unusual $\ln \rs$ terms in the high-density perturbative expansion, highlighting the nonanalytic nature of the correlation energy (see below).

The large-$\rs$ (or low-density) expansion reads \cite{Fuc-PRSL-35,Car-PR-61,CarColFei-PR-61,LooGil-WIREs-16}
\begin{equation}
	\titou{\eps_\text{xc}(\rs,\zeta) = \frac{\eta_0}{\rs} + \frac{\eta_1}{\rs^{3/2}} + \frac{\eta_2(\zeta)}{\rs^2} + \cdots}
\end{equation}
\titou{The first two terms, $\eta_0$ and $\eta_1$, are} assumed to be strictly independent of the spin polarization due to the short-range nature of spin interactions.
In this strong-coupling (or strictly-correlated) regime, the potential energy dominates over the kinetic energy, causing the electrons to localize at lattice points that minimize their (classical) Coulomb repulsion. 
These minimum-energy configurations are known as Wigner crystals. \cite{Wig-PR-34}

\section{Excited-State UEGs}
\label{sec:ES}

\begin{figure}
  \centering
  \includegraphics[width=\linewidth]{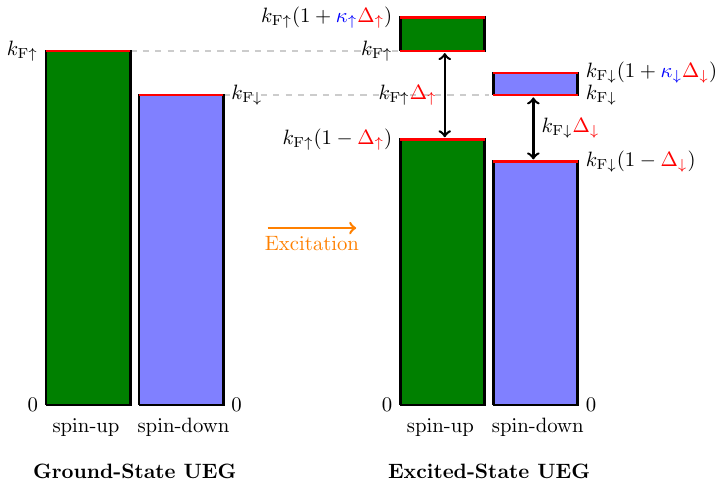}
  \caption{
  Schematic representation of ground- and excited-state UEGs.
  In a ground-state UEG (left), all electronic levels are filled from $k=0$ to the Fermi level at $k = \kFup$ for the spin-up electrons and  $k = \kFdw$ for the spin-down electrons.
  In an excited-state UEG, a gap of magnitude $\kFs\Dels$ opens at the Fermi level for each spin manifold with $0 \le \Dels \le 1$.
  The electrons in the energy levels from $\kFs(1-\Dels)$ to $\kFs$ are excited to occupy the energy 
  levels from $\kFs$ to $\kFs(1+\kaps\Dels)$. 
  The factor $0 \le \kaps \le 1$ is determined such that the spin-$\sigma$ density of the ground- and excited-state UEGs are identical.
  The red lines indicate regions where the infrared catastrophe may occur due to the vanishing gap between occupied and unoccupied states.}
   \label{fig:UEGs}
\end{figure}

Our model for the excited states of the UEG is depicted in Fig.~\ref{fig:UEGs}.
In an excited-state UEG, a gap of magnitude $\kFs\Dels$ opens at the Fermi level for each spin manifold (with $0 \le \Dels \le 1$).
The electrons in the energy levels from $\kFs(1-\Dels)$ to $\kFs$ are excited to occupy the energy levels from $\kFs$ to $\kFs(1+\kaps\Dels)$.
The parameter $0 \le \kaps \le 1$ is determined such that the spin-$\sigma$ density of the ground- and excited-state UEGs are identical (see below).
From this model, one can easily recover the ground state by setting $\Dels = 0$.
Note that the excitation process is independent for each spin channel.
A priori, this model is designed to model spin-allowed transitions.
Spin-forbidden transitions, such as singlet-triplet excitations, may require a generalization of the present model.

For the spin-$\sigma$ electrons, the occupation is
\begin{equation}
	f_{k\sigma} =
	\begin{cases}
		1	&	0 \le k \le \kFs(1-\Dels)
		\\
		0	&	\kFs(1-\Dels) < k < \kFs
		\\
		1	&	\kFs \le k \le \kFs(1+\kaps\Dels)
		\\
		0	&	k < \kFs(1+\kaps\Dels)
	\end{cases}
\end{equation}
The parameter $\kaps$ is necessary as the density of states increases with $k$.
(In other words, large values of $k$ accommodate more electrons than smaller values of $k$.)

\begin{figure}
  \centering
  \includegraphics[height=0.48\linewidth]{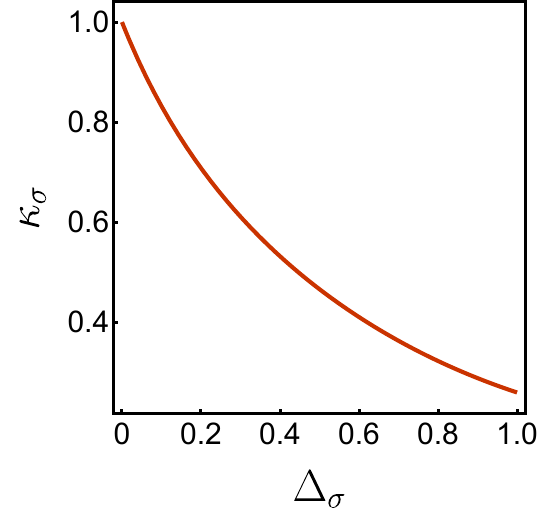}
  \includegraphics[height=0.48\linewidth]{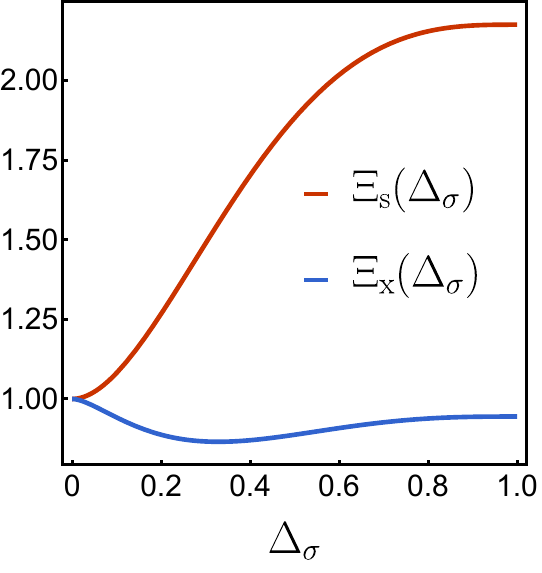}
  \caption{Left: $\kaps$ as a function of $\Dels$, as given by Eq.~\eqref{eq:kap_vs_sig}.
  Right: $\Xi_\text{s}$ and $\Xi_\text{x}$ as functions of $\Dels$, as given by Eqs.~\eqref{eq:Xis_vs_sig} and \eqref{eq:Xix_vs_sig}, respectively.}
   \label{fig:vs_Del}
\end{figure}

Such an excited-state UEG has a uniform spin density
\begin{equation}
\begin{split}
	\rho_\sigma
	& = \int_{0}^{\infty} f_{k\sigma} \frac{k^2}{2\pi^2}\dd{k} 
	\\
	& = \int_{0}^{\kFs(1-\Dels)} \frac{k^2}{2\pi^2}\dd{k} + \int_{\kFs}^{\kFs(1+\kaps\Dels)} \frac{k^2}{2\pi^2}\dd{k} 
	\\
	& = \qty[1 - 3\Dels(1-\kaps) + 3\Dels^2 (1+\kaps^2) - \Dels^3 (1-\kaps^3)] \frac{\kFs^3}{6\pi^2} 
\end{split}
\end{equation}
If one matches the density of the ground-state UEG and the excited-state UEG, one must set $\rhos = \kFs^3/(6\pi^2)$, which yields
\begin{equation} \label{eq:kap_vs_sig}
	\kaps = \frac{(1 + 3\Dels - 3\Dels^2 + \Dels^3)^{1/3} - 1}{\Dels}
\end{equation}
and has the following limits
\begin{subequations}
\begin{align}
	\lim_{\Dels\to0} \kaps & = 1
	\\
	\lim_{\Dels\to1} \kaps & = 2^{1/3} - 1 \approx 0.259921
\end{align}
\end{subequations}
The evolution of $\kaps$ as a function of $\Dels$ is represented in Fig.~\ref{fig:vs_Del}.
Note that because the excited-state density is equal to the ground-state density, the Hartree contribution is properly canceled out by the uniform positive background.

Let us now derive the reduced kinetic and exchange energies for these excited-state UEGs.
The reduced kinetic energy associated with the spin-$\sigma$ electrons is
\begin{equation}
\begin{split}
	t_{\text{s}\sigma}(\rhos,\Dels)
	& = \frac{1}{\rhos} \int_{0}^{\infty} f_k \frac{k^2}{2} \frac{k^2}{2\pi^2}\dd{k}
	\\
	& = \Xi_\text{s}(\Dels) \frac{3\kFs^2}{10}
	= \Xi_\text{s}(\Dels) C_{\text{F}} \rhos^{2/3}
\end{split}
\end{equation}
and the gap-dependent function
\begin{equation} \label{eq:Xis_vs_sig}
	\Xi_\text{s}(\Dels) = (1-\Dels)^5 + (1+\Dels\kaps)^5 - 1
\end{equation}
is depicted in Fig.~\ref{fig:vs_Del} and has the following limiting values 
\begin{subequations}
\begin{align}
	\lim_{\Dels\to0} \Xi_\text{s}(\Dels) & = 1
	\\
	\lim_{\Dels\to1} \Xi_\text{s}(\Dels) & = 2^{5/3} - 1 \approx 2.1748
\end{align}
\end{subequations}

\begin{figure}
  \centering
  \includegraphics[width=\linewidth]{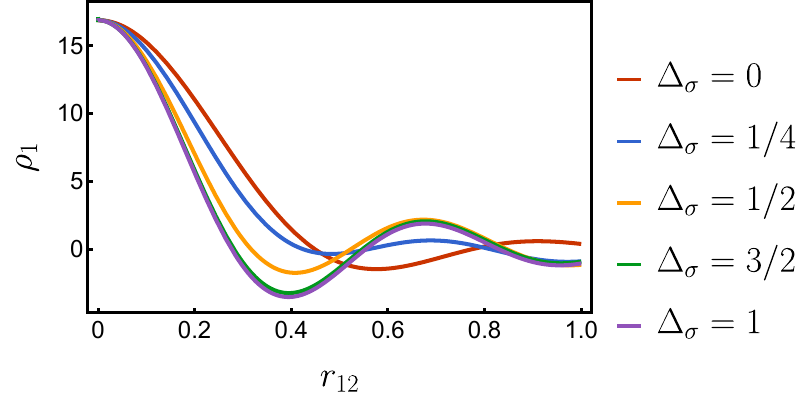}
  \caption{$\rho_1$ as a function of $r_{12}$, as given by Eqs.~\eqref{eq:rho1}, for $\kFs = 10$ and various values of $\Dels$.}
   \label{fig:rho1_vs_r12}
\end{figure}

For the exchange, we have
\begin{equation}
\begin{split}
	\eps_{\text{x}\sigma}(\rhos,\Dels)
	& = \frac{1}{2} \iint \frac{\rho_\text{x}(\br_1,\br_2)}{\abs{\br_1 - \br_2}} \dd{\br_1} \dd{\br_2}
	\\
	& = \Xi_\text{x}(\Dels) \frac{3\kFs}{4\pi}
	= \Xi_\text{x}(\Dels) C_{\text{x}} \rhos^{1/3}
\end{split}
\end{equation}
with the following gap-dependent Dirac coefficient (see Fig.~\ref{fig:vs_Del})
\begin{equation} \label{eq:Xix_vs_sig}
\begin{split} 
	\Xi_{\text{x}}(\Dels) 
	& = (1-\Dels)^4 + 4 \Dels \kaps (1+\Dels^2 \kaps^2) 
	\\
	& + 8 \Dels^2 \kappa^2 \ln 2 - \Dels^4 \kaps^4 
	\\
	& + 2 \Dels^2 \kappa^2 \Bigg[
	\qty(1 - \frac{\Dels \kaps}{2})^2 \ln (1 - \frac{\Dels \kaps}{2})
	\\
	& + 2 \qty(1 - \frac{\Dels^2 \kaps^2}{4}) \ln \qty( \frac{\Dels \kaps}{2})
	\\
	& + \qty(1 + \frac{\Dels \kaps}{2})^2 \ln (1 + \frac{\Dels \kaps}{2})
	\Bigg]
\end{split}
\end{equation}
which admits the following limiting values 
\begin{subequations}
\begin{align}
	\lim_{\Dels\to0} \Xi_\text{x}(\Dels) & = 1
	\\
	\lim_{\Dels\to1} \Xi_\text{x}(\Dels) & \approx 0.944717
\end{align}
\end{subequations}
In contrast to $\Xi_\text{s}$, $\Xi_\text{x}$ exhibits non-monotonic behavior. 
Initially, it decreases for small gap values, reaching a minimum of $0.865535$ at $\Dels \approx 0.328476$, before eventually increasing up to $\Dels = 1$.
The exchange hole of excited-state UEGs remains normalized and can be easily derived using the following expression for the one-electron reduced density matrix (see Fig.~\ref{fig:rho1_vs_r12})
\begin{multline} \label{eq:rho1}
	\rho_1(\br_1,\br_2) 
	= j_{\kFs(1+\kaps\Dels)}(r_{12}) 
	\\
	- j_{\kFs}(r_{12}) + j_{\kFs(1-\Dels)}(r_{12})
\end{multline}
As readily seen in Eq.~\eqref{eq:rho1}, the exchange hole of excited-state UEGs is a simple combination of ground-state exchange holes with different values of the Fermi wave vector.
Figure \ref{fig:rho1_vs_r12} evidences that the Fermi hole becomes tighter as $\Dels$ increases.

Now, let us study the reduced correlation energy of these excited-state UEGs.
Due to the long-range nature of the Coulomb interaction and the neglect of the kinetic energy term, the low-density limit is probably similar to the ground state (at least for the leading order proportional to $\rs^{-1}$).
This remains to be confirmed.

\titou{In Ref.~\onlinecite{GouKooGorPit-PRL-23}, it has been shown that, in the low-density limit of any finite system, any dependence on ensemble properties must vanish. Consequently, all excited-state properties become degenerate at both leading and subleading orders. Building on this, Ref.~\onlinecite{GouPit-PRX-24} presents a collection of arguments suggesting that this result very likely extends to the thermodynamic limit of UEGs.
Therefore, the same thermodynamic assumptions are expected to hold for excited UEGs, such as the one considered in this work.}
From hereon, we focus on the high-density limit and assume a non-polarized gas for the sake of simplicity.

Rayleigh-Schr\"odinger perturbation theory tells us that the second-order contribution can be decomposed as $\eps^{(2)} = \eps^{(2\text{d})} + \eps^{(2\text{x})}$,
where, for the ground state, the exchange term, $\eps^{(2\text{x})}$, is known to be finite \cite{GelBru-PR-57,OnsMitSte-AP-66} and the direct term, $\eps^{(2\text{d})}$, 
has been first shown to diverge logarithmically as $\lambda_0 \ln \rs$ for small $\rs$ by Macke \cite{Mac-ZNA-50} with $\lambda_0 = (1-\ln 2)/\pi^2 \approx 0.0310907$.

More explicitly, the direct component reads \cite{RaimesBook}
\begin{equation} \label{eq:e2d}
	\eps^{(2\text{d})} = -\frac{3}{16\pi^5} \int \frac{\dd{\bk}}{k^4} \int \dd{\bq} \int \frac{\dd{\bp}}{\bk \cdot (\bp - \bq + \bk)}
\end{equation}
where, in the case of the ground state,  $p < 1$, $q <  1$, $1 < \abs{\bp + \bk}$, and $1 < \abs{\bq - \bk}$ in units of $\kF$.
The divergent behavior can be derived by realizing that the main contribution to the integral comes from the small momentum transfer (i.e., $k \approx 0$) or, in other words, from the excitations near the Fermi surface where the energy gap between occupied and vacant states vanishes (infrared catastrophe).
This issue is particularly problematic in coupled-cluster theory, where perturbative corrections systematically exhibit infrared divergences. \cite{MasIrmSchGru-PRL-23,NeuBer-PRL-23}

By assuming that $p \approx 1$ and $q \approx 1$ and neglecting the terms proportional to $k^2$, one can show that $1-kx \le p \le 1$ and $1-ky \le q \le 1$, where $\bk \cdot \bp = kpx$ and $\bk \cdot \bq =  - kqy$. This leads to
\begin{multline}
	\int \dd{\bq} \int \frac{\dd{\bp}}{\bk \cdot (\bp - \bq + \bk)} 
	\\
	 \approx (2\pi)^2 \int_0^1 \dd{x} \int_0^1 \dd{y} \int_{1-kx}^{1} \dd{p} \int_{1-ky}^{1} \frac{\dd{q}}{k(x+y)}
\end{multline}
Finally, the integration over $k$ in the range $\sqrt{\rs} < k  < 1$ (where the upper limit is arbitrary and $\sqrt{\rs}$ corresponds to the characteristic wave vector below which the Coulomb interaction is effectively screened \cite{GelBru-PR-57}) yields
\begin{equation}
\begin{split}
	\eps^{(2\text{d})} 
	& \approx -\frac{3}{16\pi^5} \int_{\sqrt{\rs}}^{1} \frac{4\pi k^2\dd{k}}{k^4} (2\pi)^2 \frac{2k}{3} (1-\ln 2)
	\\
	& \sim \frac{1-\ln 2}{\pi^2} \ln \rs
\end{split}
\end{equation}

\begin{figure}
  \centering
  \includegraphics[width=0.6\linewidth]{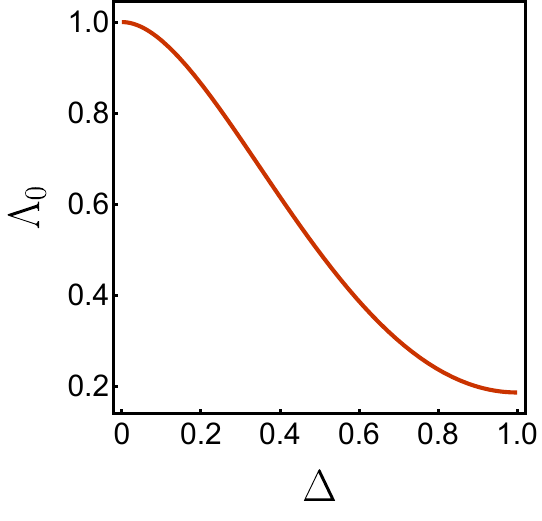}
  \caption{$\Lambda_0$ as a function of $\Delta$, as given by Eq.~\eqref{eq:lam0_vs_Del}.}
   \label{fig:Lam_vs_Del}
\end{figure}

For the excited-state UEGs, the infrared divergence originates from various (de)excitation processes in which the energy difference between occupied and vacant states vanishes (see red lines in Fig.~\ref{fig:UEGs}).
By considering the various admissible regions of $\bp$, $\bq$, and $\bk$ in Eq.~\eqref{eq:e2d}, we have identified six divergent terms:
\begin{itemize}
\item Excitations from occupied states with $0 < p < 1-\Delta$ and $0 < q < 1-\Delta$ to unoccupied states with $1-\Delta < \abs{\bp + \bk} < 1$ and $1-\Delta < \abs{\bq - \bk} < 1$;
\item De-excitations from occupied states with $1 < p < 1+\kappa\Delta$ and $1 < q < 1+\kappa\Delta$ to unoccupied states with $1-\Delta < \abs{\bp + \bk} < 1$ and $1-\Delta < \abs{\bq - \bk} < 1$;
\item Excitations from occupied states with $1 < p < 1+\kappa\Delta$ and $1 < q < 1+\kappa\Delta$ to unoccupied states with $1+\kappa\Delta < \abs{\bp + \bk}$ and $1+\kappa\Delta < \abs{\bq - \bk}$;
\item Mixed process corresponding to excitations from occupied states with $0 < p < 1-\Delta$ to unoccupied states with $1-\Delta < \abs{\bp + \bk} < 1$, combined with de-excitations from occupied states with $1 < q < 1+\kappa\Delta$ to unoccupied states with $1-\Delta < \abs{\bq - \bk} < 1$;
\item Mixed process corresponding to de-excitations from occupied states with $1 < p < 1+\kappa\Delta$ to unoccupied states with $1-\Delta < \abs{\bp + \bk} < 1$, combined with excitations from occupied states with $1 < q < 1+\kappa\Delta$ to unoccupied states with $1+\kappa\Delta < \abs{\bq - \bk}$;
\item Excitations from occupied states with $0 < p < 1-\Delta$ to unoccupied states with $1-\Delta < \abs{\bp + \bk} < 1$, combined with excitations from occupied states with $1 < q < 1+\kappa\Delta$ to unoccupied states with $1+\kappa\Delta < \abs{\bq - \bk}$;
\end{itemize}
Each of these processes leads to a logarithmic divergence of the correlation energy in the high-density limit, with a distinct dependence on $\Delta$.
\titou{For excited-state UEGs, the $\Delta$-dependent direct component has the following form
\begin{equation}
	\eps^{(2\text{d})}(\Delta) \sim \lambda_0(\Delta) \ln \rs
\end{equation}
and,} following a similar procedure as for the ground state (see above), one can show that the corresponding gap-dependent coefficient can be decomposed as
\begin{equation} \label{eq:lam0_vs_Del}
	\lambda_0(\Delta) =  \Lambda_0(\Delta) \lambda_0 = \frac{1}{\pi^2} \sum_{k=1}^6 \lambda_{0}^{(k)}
\end{equation}
with
\begin{align*}
	\lambda_0^{(1)} & = (1-\Delta)^3 F(1,1)
	&
	\lambda_0^{(2)} & = F(1,1)
	\\
	\lambda_0^{(3)} & = (1+\kappa\Delta)^3 F(1,1)
	&
	\lambda_0^{(4)} & = - 2 F(1-\Delta,1)
	\\
	\lambda_0^{(5)} & = - 2 F(1,1+\kappa\Delta)
	&
	\lambda_0^{(6)} & = 2 F(1-\Delta,1+\kappa\Delta)
\end{align*}
and
\begin{multline}
	F(\alpha,\beta) = \alpha^2\beta + \alpha\beta^2 + \alpha^3 \ln \alpha + \beta^3 \ln \beta
	\\
	- (\alpha^3 + \beta^3) \ln (\alpha + \beta) 
\end{multline}
In the limit of a vanishing gap, the ground-state behavior is recovered, i.e., $\lambda_0(\Delta = 0) = \lambda_0 = (1-\ln2)/\pi^2$.
As $\Delta$ approaches 1, the behavior remains logarithmic, though weaker, with $\lambda_0(\Delta = 1) \approx 0.00578826$.
The evolution of $\Lambda_0(\Delta) = \lambda_0(\Delta)/\lambda_0$ as a function of $\Delta$ is shown in Fig.~\ref{fig:Lam_vs_Del}.
Higher-order terms in $\rs$ have yet to be explored.
This is left for future work.

\section{Conclusion}
\label{sec:conclusion}
To conclude, \titou{this} work introduces a generalization of the UEG paradigm for excited states \titou{and reports} closed-form expressions for the reduced kinetic and exchange energies of these excited-state UEGs as functions of the density and the gap. 
Additionally, the leading-order term of the correlation energy in the high-density limit \titou{is derived}.
By modifying the UEG to include a gap at the Fermi surface arising from excitations near the Fermi level, \titou{this new paradigm provides} a foundation for constructing state-specific functionals beyond ensemble-based approaches. 
This pure-state approach represents a significant step toward developing (semi)local approximations capable of accurately describing excited-state properties within DFT. 

\titou{One might reasonably argue that the UEG bears little resemblance to a real-life molecule. 
Nevertheless, the LDA, originally derived from the UEG, has proven surprisingly successful in modeling the electronic structure of complex materials and molecular systems, or at the very least, in serving as a foundation for more sophisticated approximations.
We hope that this transferability from the UEG to molecular systems extends to excited states as well, though this remains to be thoroughly validated. 
It is important to recognize that, in practice, a global property of the UEG, such as the exchange or correlation energy computed over the entire system (i.e., involving all electrons), is mapped onto a local property within a molecule, such as the exchange or correlation energy at a specific grid point. This same philosophy underlies the construction of our excited-state UEG models.
In the present approach, electrons are considered to be excited from a fraction of the occupied band into a fraction of the unoccupied band, a global property of the UEG, which we aim to map onto a local measure that reflects the degree of excitation at each point in a molecular system.
That said, we acknowledge that this mapping and its implications require further investigation, and we hope to address some of these open questions in future work.}

In terms of perspectives, (semi)local functionals based on excited-state UEGs \titou{are currently being developed} by embedding an additional variable that quantifies the degree of excitation directly into the functional. 
This approach opens new avenues for improving density-functional approximations beyond ground-state DFT. 
Future work will focus on refining this framework and evaluating its performance in practical applications.

\acknowledgements{
The author would like to thank Tim Gould, Andreas Savin, Hugh Burton, and Andr\'e Laestadius for insightful discussions.
This project has received funding from the European Research Council (ERC) under the European Union's Horizon 2020 research and innovation programme (Grant agreement No.~863481).}

\section*{Data availability statement}
Data sharing is not applicable to this article as no new data were created or analyzed in this study.

\section*{References}

\bibliography{biblio}

\end{document}